\documentstyle[aps,epsf,psfrag,float,amssymb,amstex]{revtex}

\begin{document}
\draft
\title{Thermodynamics of Reissner-Nordstr\"om-anti-de~Sitter black
  holes in the grand canonical ensemble}
\author{Claudia S. Pe\c ca} 
\address{Departamento de F\' \i sica, Instituto Superior T\'ecnico,
  Av. Rovisco Pais 1, 1096 Lisboa Codex, Portugal} 
\author{Jos\'e P. S. Lemos}
\address{Departamento de Astrof\' \i sica. Observat\'orio
  Nacional-CNPq, Rua General Jos\'e Cristino 77, 20921 Rio de Janeiro,
  Brazil \\ and 
Departamento de F\' \i sica, Instituto Superior T\'ecnico,
  Av. Rovisco Pais 1, 1096 Lisboa Codex, Portugal} 
\date{\today}
\maketitle
\begin{abstract}
The thermodynamical properties of the
Reissner-Nordstr\"om-anti-de~Sitter black hole in the grand canonical
ensemble are investigated using York's formalism. The black hole is
enclosed in a cavity with finite radius where the temperature and
electrostatic potential are fixed. The boundary conditions allow one to
compute the relevant thermodynamical quantities, e.g. thermal energy,
entropy and charge. The stability conditions imply that there are
thermodynamically stable black hole solutions, under certain
conditions. By taking the boundary to infinity, and leaving the event
horizon and charge of the black hole fixed, one rederives the
Hawking-Page action and Hawking-Page specific heat.  Instantons with
negative heat capacity are also found.
\end{abstract}

\section{Introduction}

The path-integral approach to the thermodynamics of black holes was
originally developed by Hawking {\it et al.}
\cite{Hartle,GibH,Hawklect}. In this approach the thermodynamical
partition function is computed from the path-integral in the
saddle-point approximation, thus obtaining the thermodynamical laws
for black holes. 

In the path-integral approach we can use the three different ensembles:
microcanonical, canonical and grand canonical. Due to difficulties
related to  stability of the black hole in the canonical ensemble, the
microcanonical ensemble was originally considered
\cite{Hawklect,Hawk76}. However, further developments by York
{\it et al.} \cite{York,WhitYork,Whit90,grqcbrown} allowed to
define the canonical ensemble.
Effectively, by carefully defining the boundary conditions, one can
 obtain the partition function of a black hole in thermodynamical
equilibrium. This approach was further developed to include other
ensembles \cite{Brown90}, and to study  charged black holes in the
grand canonical ensemble \cite{Braden} and black holes in asymptotically
anti-de~Sitter spacetimes \cite{Brown94,jolien,Louko}. This approach
was also applied to black holes in two  \cite{Lemos} and three
\cite{Brown94} dimensions.  

In York's formalism the black hole is enclosed in a cavity with a
finite radius. The boundary conditions are defined according to
the thermodynamical ensemble under study. 
Given the boundary conditions and imposing the appropriate constraints,
one can compute a reduced action suitable for doing black hole
thermodynamics \cite{Braden,York89}. Evaluating  this reduced action
at its stable stationary point one obtains the corresponding classical
action, which is related to a thermodynamical potential. 
In the canonical ensemble this thermodynamical potential corresponds
to the Helmholtz free energy, while for the grand canonical ensemble
the thermodynamical potential is the grand canonical
potential \cite{GibH,Braden}. From the thermodynamical potential one
can compute all the relevant thermodynamical quantities and relations
\cite{Callen}.  

Some controversy has appeared related to the boundary conditions
chosen in this formalism \cite{Brown94,HP83,Page}. More precisely, Hawking 
and Page \cite{HP83,Page} fix the Hawking temperature of the black
hole (i.e. the temperature defined so that the respective Euclidean
metric has no conical singularity at the horizon), while
York {\it et al.} \cite{York,Braden,Brown94} fix the local temperature
at a finite radius, where the boundary conditions are defined. For
asymptotically flat spacetimes the two formalism 
coincide, because at infinity the local temperature is equal to the
Hawking temperature. On the contrary, for asymptotically anti-de~Sitter
spacetimes the two procedures disagree,  since the local temperature is
redshifted to zero at infinity and is equal to the Hawking
temperature only in the region where spacetime has a flat metric. Louko and
Winters-Hilt \cite{Louko} have studied the thermodynamics of the
Reissner-Nordstr\"om-anti-de~Sitter black hole fixing a
renormalized temperature at infinity that corresponds to the same
procedure used in \cite{HP83,Page}. In this paper we have chosen to
follow York's formalism \cite{York,Braden,Brown94} and study the
thermodynamics of the Reissner-Nordstr\"om-anti-de~Sitter black hole
fixing the local temperature at finite radius.

We find that the two procedures give some identical
results, e.g., in both procedures the Hawking-Bekenstein formula for
the entropy \cite{Hawk75,Bek73} is obtained. 
In addition, by that taking the boundary to infinity,
and leaving the event horizon and charge of the black hole fixed, 
we rederive the Hawking-Page action and Hawking-Page specific heat
from York's formalism. 
However, the value for the energy at infinity differs depending on which
procedures one uses. In \cite{Louko} it was found that the energy at
infinity is equal to the mass of the black hole, a result that does
not hold here. These results conform with the similarities and
differences found for the Schwarzschild-anti-de~Sitter black hole in
\cite{HP83,Brown94}.  

In section \ref{s:action} we briefly introduce York's formalism. In
section \ref{s:RNADS} we compute the reduced action for the
Reissner-Nordstr\"om-anti-de~Sitter black hole and evaluate its
thermodynamical quantities. In section \ref{s:sol} we analyze the
black hole solutions. In section \ref{stab} we study the local and
global stability of these solutions. The limit where the boundary is
taken to infinity mentioned above is studied in section
\ref{s:infty}. Finally some special cases are briefly referred in
section \ref{s:special}.

\section{The action}
\label{s:action}

The Euclidean Einstein-Maxwell  \cite{Hawklect} is given by
\begin{equation}
  \label{action}
  I = - \frac {1} {16 \pi} \int_{\cal M} d^4x \, \sqrt g \,(R-2
\Lambda) \  + \  \frac {1} {8 \pi} \int_{\partial {\cal M}} d^3x \,
\sqrt{h} \, K  \  - \ \frac{1}{16 \pi} \int_{\cal M} d^4 x\
\sqrt{g}\, F_{\mu\nu} F^{\mu\nu} \ -\ I_{\rm subtr} \ ,
\end{equation}
where ${\cal M}$ is a compact region with boundary $\partial {\cal
  M}$, $R$ is the scalar curvature, $\Lambda$ the cosmological
constant, $g$ the determinant of the Euclidean metrics, $K$ the trace
of the extrinsic curvature of the boundary $\partial {\cal M}$, $h$
is the determinant of the Euclidean induced metrics on the boundary,
$F_{\mu\nu}=\partial_\mu A_\nu-\partial_\nu A_\mu$ is the Faraday
tensor and $I_{\rm subtr}$ is an arbitrary term that can be used to
define the zero of the energy as will be seen later. 

In order to set the nomenclature we follow \cite{Braden} in this
section. 
We consider a spherical symmetric static metric of the form \cite{Braden}
\begin{equation}
  \label{metric}
  ds^2=b^2 d\tau^2+a^2 dy^2+r^2 d\Omega^2\ ,
\end{equation}
where $a$, $b$ and $r$ are functions only of the radial coordinate
$y\in[0,1]$. The Euclidean time $\tau$ has period $2 \pi$.  The event 
horizon, given by $y=0$, has radius $r_+=r(0)$ and area $A_+=4 \pi
r_+^2$. The boundary is given by $y=1$ and at this boundary  the
thermodynamical variables  defining the ensemble are fixed. The
boundary is a two-sphere with area 
$A_B=4 \pi r_B^2$, where $r_B=r(1)$. We will consider the grand canonical
ensemble, where heat and charge can flow in and out through the boundary to
maintain a constant temperature $T\equiv T(r_B)$ and electrostatic 
potential $\phi \equiv \phi(r_B)$ at the boundary.   

We impose a black hole topology to the metric (\ref{metric}), by using
the  conditions,  $b(0)=0$, $\left.\frac{b'}{a} \right|_{y=0} = 1$ and
$\left. \left(\frac{r'}{a} \right)^2 \right|_{y=0} = 0$, see
\cite{Braden}. 

Evaluating the action (\ref{action}) for the metric (\ref{metric}), 
\begin{eqnarray}
  \label{Ig}
  I &=& \frac{1}{2} \int_0^{2\pi}d\tau \int_0^{1}dy \  \left(-2
    \frac{r\, b'\, r'}{a}-\frac{b\, {r'}^2}{a}-a\, b+\Lambda\, a\, b\,
    r^2 \right)- \frac{1}{2} \int_0^{2\pi}d\tau \  \left. \frac{(b\,
      r^2)'}{a} \right|_{y=0}\ - \nonumber \\ &{}&- \frac{1}{2}
  \int_0^{2 \pi}d\tau\ \int_0^1 dy \ \frac{r^2}{a\,b} \ {A'_\tau}^2 -\
  I_{\rm subtr}\ . 
\end{eqnarray}

In order to obtain the reduced action one uses the proper constraints. 
For the gravitational part of the action $I_g$ given in (\ref{Ig}), the
constraint used is the Hamiltonian constraint \cite{Braden,York89}
\begin{equation}
  \label{hamconstr}
  {G^\tau}_\tau + \Lambda {g^\tau}_\tau = 8 \pi {T^\tau}_\tau \ .
\end{equation}
In addition, for the matter fields part of the action we use  Maxwell
equations ${F^{\mu\nu}}_{;\nu} = 0$.

The thermodynamical quantities and relations are obtained from the
``classical action'' $\tilde I$ (defined as the reduced action  evaluated at
its locally stable stationary points) using the well known relation
between the ``classical action'' and the thermodynamical potential 
\begin{equation}
  \label{IbetaF}
  \tilde I = \beta F \ .
\end{equation}
Here $F$ is the grand canonical potential since we are considering
the grand canonical ensemble. All the thermodynamical quantities can
be obtained from $F$ using the classical thermodynamical relations
(see for example \cite{Callen}).  

\section{The Reissner-Nordstr\"om-anti-de~Sitter black hole}
\label{s:RNADS}

The Reissner-Nordstr\"om-anti-de~Sitter black hole in the grand
canonical ensemble is obtained using a negative cosmological constant
$\Lambda$ and the boundary conditions $T\equiv T(R_B)$ and $\phi\equiv
\phi(r_B)$, where $r_B$ is the boundary radius of the spherical
cavity, $T$ the temperature at the boundary and $\phi$ the
electrostatic potential difference between the horizon and the
boundary. Instead of $T$ we can also use its inverse $\beta$. 

The reduced action for Reissner-Nordstr\"om-anti-de~Sitter black hole
is obtained from the Euclidean Einstein-Hilbert-Maxwell action $I$ given
in (\ref{action}). 
For simplicity we split the action in two terms
$I=I_g+I_m$, where $I_g$ is the gravitational term
and $I_m$ the matter field term. To obtain the
reduced action we use the Hamiltonian constraint (\ref{hamconstr}) 
and the Maxwell equations. 

The evaluation of $I_m$ is identical to the case $\Lambda=0$ and can
therefore be found in \cite{Braden}, 
\begin{equation}
  \label{IaRNADS}
  I_m= -\frac{1}{2}\, \beta\, e \, \phi\ ,
\end{equation}
where $e$ is the electrical charge of the black hole and $\phi$ is the
difference of potential between $y=0$ and $y=1$. 
To evaluate the gravitational term $I_g$ (\ref{Ig}) we use as
mentioned above the constraint (\ref{hamconstr})
\begin{equation}
  \label{hc}
  {G^\tau}_\tau + \Lambda\, {g^\tau}_\tau = 8 \pi\, {T^\tau}_\tau\ .
\end{equation}
The component of the Einstein tensor ${G^\tau}_\tau$ for the metric
(\ref{metric}) is
\begin{equation}
  \label{G}
     {G^\tau}_\tau=\frac{{r'}^2}{a^2 r^2}-\frac{1}{r^2}+\frac{2 {r'}'}{a^2
    r}-\frac{2 a' r'}{a^3 r}\ .
\end{equation}
The stress-energy tensor component ${T^\tau}_\tau$  is given by
\begin{equation}
  \label{T}
  {T^\tau}_\tau=\frac{1}{8\, \pi} \left(\frac{{A_\tau}'}{a\, b}
  \right)^2 =-\frac{1}{8\pi}\frac{e^2}{r^4}\ .
\end{equation}
Substituting (\ref{T}) and (\ref{G}) in (\ref{hc}) we obtain
\begin{equation}
  \label{Lambda}
    \Lambda= \left(\frac{{A_\tau}'}{a\, b}
  \right)^2 -\frac{{r'}^2}{a^2 r^2} +\frac{1}{r^2} -\frac{2 {r'}'}{a^2
    r}+\frac{2 a' r'}{a^3 r}. 
\end{equation}
Rearranging terms in equation (\ref{Lambda}) and using (\ref{T}) we
obtain 
\begin{equation}
  \label{ch1}
    \frac{1}{r^2\,r'} \left\{ \left[ r \left( \frac{{r'}^2}{a^2}-1 \right)
      \right]'+ \frac{e^2\,r'}{r^2}+\Lambda\,r^2\,r'\right\}=0 \ .
\end{equation}
Integrating and simplifying the previous equation yields
\begin{equation}
  \label{r'a}
    \left(\frac{r'}{a}\right)^2=1-\frac{2\,M}{r}+\frac{e^2}{r^2}+
  \alpha^2\,r^2 \ ,
\end{equation}
where $2 M$ is an integration constant and $\alpha^2=-\Lambda/3$. The
integration constant $2 M$ can be evaluated using the black hole
topology condition $\left. \left(\frac{r'}{a}\right)^2 \right|_{y=0} =
0$,    
\begin{equation}
  \label{2M}
   2\,M=r_++\frac{e^2}{r_+}+\alpha^2\,r_+^3\ .
\end{equation}
This is the known relation between the ADM mass of the
Reissner-Nordstr\"om-anti-de~Sitter black hole and its event horizon
radius. 

Substituting (\ref{Lambda}) in (\ref{Ig}) yields
\begin{eqnarray}
  \label{IgRNADS.}
    I_g^*&=&\frac{1}{2} \int_0^{2\pi}d\tau \int_0^{1}dy \  \left(-2 \frac{r\,
      b'\, r'}{a}-2 \frac{b\, {r'}^2}{a}-2 \frac{ b\,r\,{r'}'}{a} +2
    \frac{b\,r\,a' r'}{a^2}-\ \frac{r^2}{a\,b} \ {A'_\tau}^2
  \right)-{} \nonumber \\ &{}&-\frac{1}{2} \int_0^{2\pi}d\tau
  \  \left. \frac{(b\, r^2)'}{a} \right|_{y=0}\  -\
  I_{\rm subtr}\nonumber \\ 
&=&-\int_0^{2\pi}d\tau \int_0^{1}dy \ \left(\frac{b\,r\,r'}{a}\right)'
\ -\frac{1}{2} \beta \, e \, \phi \ 
-\frac{1}{2} \int_0^{2\pi}d\tau \  \left. \frac{b'\, r^2}{a}
\right|_{y=0}\  -\ I_{\rm subtr} \ ,
\end{eqnarray}
where we have used the topology conditions given in section
\ref{s:action} and to evaluate the term in $A_\tau$, we used
(\ref{IaRNADS}), since this term is identical to $I_m$.  

The first term after the second equality in equation (\ref{IgRNADS.})
can be evaluated by integrating and substituting equations (\ref{r'a})
and (\ref{2M}). The respective third term is integrated and using the
topology conditions  gives
$-\pi r_+^2$.  Following this procedure, we obtain
\begin{equation}
  \label{IgRNADS}
    I_g^*=-\beta\, r_B\, f\left( r_B; \, r_+,\, e ,\, \alpha \right)\ 
  -\frac{1}{2}\, \beta \,e\,\phi-\pi r_+^2 -I_{\rm subtr} \ ,
\end{equation}
where the inverse temperature at the boundary $\beta$ is given
by the proper length of the time coordinate at the boundary,  
$\beta \equiv T^{-1} = \int_{0}^{2 \pi} b(1) \, d\tau = 2 \, \pi \,
b(1)$ and 
\begin{equation}
\label{f}
f\left( r_B; \, r_+,\, e ,\, \alpha \right)=
 \sqrt{1-\frac{r_+}{r_B}-\frac{e^2}{r_+\,r_B}-\alpha^2\,
  \frac{r_+^3}{r_B}+\frac{e^2}{r_B^2}+\alpha^2\,r_B^2} \ . 
\end{equation}

Adding (\ref{IaRNADS}) and (\ref{IgRNADS}), yields the reduced action
\begin{equation}
\label{Irnads}
I^*=-\beta\, r_B\, f\left( r_B; \, r_+,\, e ,\, \alpha \right)\ 
  - \beta \,e\,\phi-\pi r_+^2 -I_{\rm subtr}\ .
\end{equation}
The term $I_{\rm subtr}$ is of the form $\beta E_{\rm subtr}$, where
$E_{\rm subtr}$ is a constant that does not depend on $\beta$ or
$\phi$, since $I_{\rm subtr}$ is an arbitrary term that can be used to
fix the zero of the energy but cannot affect other thermodynamical
variables \cite{York}. For convenience, we use for the zero of the energy 
\begin{equation}
\label{E0rnads}
E_{\rm ADS}=E(r_+=0,e=0)=0\ ,
\end{equation}
where $E_{\rm ADS}$ is the thermal energy of anti-de~Sitter
spacetime. 

To evaluate  $I_{\rm subtr}$, we compute the thermal energy of the
Reissner-Nordstr\"om-anti-de~Sitter black hole from (\ref{Irnads}) and
use condition (\ref{E0rnads}). The thermal energy is given by
\cite{Callen} 
\begin{eqnarray}
\label{Ernads}
E &=& F+ \beta \left( \frac{\partial F}{\partial
      \beta}\right)_{\phi,r_B} - \left( \frac{\partial F}{\partial
      \phi} \right)_{\beta,r_B} \phi  =
\left( \frac{\partial \tilde I}{\partial \beta}\right)_{\phi,r_B}-
\frac{\phi}{\beta} \left( \frac{\partial \tilde I}{\partial \phi}
\right)_{\beta,r_B} = \nonumber \\ &=&-r_B\, f\left( r_B; \, r_+,\, e
  ,\, \alpha \right) \ - E_{\rm subtr} 
\end{eqnarray}
 where $F$ is the grand canonical potential and we have used
(\ref{IbetaF}). Although the reduced action $I^*$ is not the classical
action (therefore we cannot write $I^*= \beta F$), the energy has the
form given in (\ref{Ernads}). This is because the classical action
$\tilde I$ corresponds to the minimum of the reduced action and
therefore the equalities 
$\left( \frac{\partial \tilde I}{\partial \beta}\right)_{\phi,r_B}=\left(
  \frac{\partial I^*}{\partial \beta}\right)_{\phi,r_B,r_+,e}$ and 
$\left( \frac{\partial \tilde I}{\partial \phi}\right)_{\beta,r_B}=\left(
  \frac{\partial I^*}{\partial \phi}\right)_{\beta,r_B,r_+,e}$
 hold. However, $r_+$ and $e$ in (\ref{Ernads}) are not  free
 parameters, they depend on the boundary conditions (i.e. on the
 values of $\beta$, $\phi$ and $r_B$) and on the cosmological
 constant. The functions $r_+=r_+(\beta,\phi,r_B,\alpha)$  and
 $e=e(\beta,\phi,r_B,\alpha)$ are obtained from the equilibrium 
conditions $\frac{\partial I^*}{\partial r_+}=0$ and $\frac{\partial
  I^*}{\partial e}=0$ as will be seen later. 

Using equation (\ref{E0rnads}) on (\ref{Ernads}), yields
\begin{equation}
\label{EsubtrRNADS}
E_{\rm subtr}=-r_B\, f_0\left( r_B;\, \alpha \right)
\end{equation}
where $f_0\left( r_B;\, \alpha \right)=f\left( r_B; \, 0,\, 0,\,
  \alpha \right)=\sqrt{1+\alpha^2\,r_B^2}$.

Substituting (\ref{EsubtrRNADS}) in (\ref{Irnads}), we finally obtain
the reduced action for the Reissner-Nordstr\"om-anti-de~Sitter black
hole 
\begin{equation}
  \label{IRNADS}
  I^*=\beta\, r_B\,\left[
f_0\left( r_B;\, \alpha \right)- f\left( r_B; \, r_+,\, e ,\, \alpha \right)
 \right]
  - \beta \,e\,\phi-\pi r_+^2 \ .
\end{equation}

Similarly substituting (\ref{EsubtrRNADS}) in (\ref{Ernads}), we
obtain its thermal energy 
\begin{equation}
\label{ERNADS}
E= r_B\,\left[
 f_0\left( r_B;\, \alpha \right)- f\left( r_B; \, r_+,\, e ,\, \alpha
 \right) \right]\ .
\end{equation}

The mean value of the charge is 
\begin{equation}
  \label{QRNADS}
  Q = -\left( \frac{\partial F}{\partial \phi} \right)_{\beta,r_B} = 
- \frac{1}{\beta} \left( \frac{\partial \tilde I}{\partial \phi}
\right)_{\beta,r_B} = e \ .
\end{equation}

The entropy is obtained from 
\begin{equation}
  \label{SRNADS}
  S = \beta^2 \left(\frac{\partial F}{\partial \beta}\right)_{\phi,r_B}=
\beta \  \left(\frac{\partial \tilde I}{\partial \beta}\right)_{\phi,r_B}
  - \tilde I = \pi r_+^2\ ,
\end{equation}
where equation (\ref{IbetaF}) was used. Since $A_+/4=\pi\, r_+^2$,
where $A_+$ is the area of the event horizon, this is the usual
Hawking-Bekenstein entropy \cite{Hawk75,Bek73}.

As mentioned above, the event horizon radius $r_+$ and electric charge
$e$ of the black hole for the given boundary conditions, i.e. $\beta$,
$\phi$ and $r_B$, are obtained by evaluating the locally stable
stationary points of the reduced action with respect to $r_+$ and $e$
\cite{Braden}. Effectively, once  the values of $\beta$, $\phi$ and
$r_B$ are held fixed by the boundary 
conditions, then the reduced action is a function of only $r_+$ and
$e$, i.e. $I^*=I^*(r_+,e)$. The local stability conditions are then
($i$) $\nabla I^*=0$ and ($ii$) the Hessian matrix is positive
definite. The latter condition corresponds to a condition of dynamical
as well as thermodynamical stability \cite{Braden} and will be
discussed in section \ref{stab}. We will start by investigating the
first condition. 

The condition of stationarity $\nabla I^*=0$ gives 
\begin{equation}
\label{betaRNADS}
\beta=
\frac{2 \pi}{\kappa} \ f\left( r_B;\, r_+,\, e ,\, \alpha \right) \ ,
\end{equation}
where $\kappa= \frac{r_+^2 - e^2 + 3\alpha^2r_+^4}{2r_+^3}$ is the
surface gravity of the horizon, and 
\begin{equation}
\label{phiRNADS}
\phi=\left(\frac{e}{r_+}-
\frac{e}{r_B}\right)\, f\left( r_B; \, r_+,\, e ,\, \alpha
\right)^{-1} \ .
\end{equation}
These are the inverse Hawking temperature and the difference in the
electrostatic potential between $r_+$ and $r_B$ blueshifted from
infinity to $r_B$, respectively. 

Inverting these two equations, $r_+$ and $e$ are obtained as functions
of the boundary conditions and the cosmological constant. We define
the more convenient variables 
\begin{equation}
\label{newRNADS}
  x\equiv \frac{r_+}{r_B}\ \ ,\ \ \ \ \ q\equiv\frac{e}{r_B}\ \ ,\ \ \
\ \ \overline{\beta}\equiv \frac{\beta}{4\pi\,r_B}\ \ ,\ \ \ \ \ 
\overline{\alpha}^2\equiv \alpha^2\, r_B^2 \ . 
\end{equation}
In these variables (\ref{betaRNADS}) and (\ref{phiRNADS}) are written 
\begin{eqnarray}
  \label{betarnads}
  \overline{\beta}&=&x \, \sqrt{1-x} \  \sqrt{1-\frac{q^2}{x}+
    \overline{\alpha}^2 \, (1+x+x^2)} \ \left(1-\frac{q^2}{x^2}+
    3\,\overline{\alpha}^2\,x^2 \right)^{-1} \ , \\ \label{phirnads}
  \phi&=&\frac{q}{x}\, \sqrt{1-x}\
  \left(1-\frac{q^2}{x}+\overline{\alpha}^2 \, (1+x+x^2)
  \right)^{-1/2} \ .
\end{eqnarray}
To invert this equations, we start by inverting equation
(\ref{phirnads}) to obtain $q$,
\begin{equation}
  \label{q(x)RNADS}
  q^2= x^2\, \phi^2\  [1+\overline{\alpha}^2\,(1+x+x^2)]\
  (1-x+x\,\phi^2)^{-1} \ .
\end{equation}
Substituting (\ref{q(x)RNADS}) in (\ref{betarnads}) and taking its
square, we obtain a 7th degree equation in $x$ with a double root
$x=1$ which is not a solution of the initial equations. Getting rid of
that root, we obtain a 5th degree equation 
\begin{eqnarray}
  \label{x^5}
&&{\overline{\beta}^2}\,\left( -1 + {\phi^2} + 
        {\overline{\alpha}^2}\,{\phi^2} \right)^2\  \ +\  \ 
  4\,{\overline{\alpha}^2}\,{\overline{\beta}^2}\,{\phi^2}\,
   \left( -1 + {\phi^2} + 
     {\overline{\alpha}^2}\,{\phi^2} \right) \,x + 
  \left( -1 - {\overline{\alpha}^2} +
    6\,{\overline{\alpha}^2}\,{\overline{\beta}^2} -  
     12\,{\overline{\alpha}^2}\,{\overline{\beta}^2}\,{\phi^2} - \right.
     \nonumber \\&&{}- \left. 
     6\,{\overline{\alpha}^4}\,{\overline{\beta}^2}\,{\phi^2} +  
     6\,{\overline{\alpha}^2}\,{\overline{\beta}^2}\,{\phi^4} +
     10\,{\overline{\alpha}^4}\,{\overline{\beta}^2}\,{\phi^4} \right) \,{x^2}
   +  \left( 1 - {\phi^2} - 
     {\overline{\alpha}^2}\,{\phi^2} - 
     12\,{\overline{\alpha}^4}\,{\overline{\beta}^2}\,{\phi^2} + 
     12\,{\overline{\alpha}^4}\,{\overline{\beta}^2}\,{\phi^4} \right) \,{x^3} 
   + \nonumber \\ &&{} +
  {\overline{\alpha}^2}\,\left(
    9\,{\overline{\alpha}^2}\,{\overline{\beta}^2} - {\phi^2} -  
     18\,{\overline{\alpha}^2}\,{\overline{\beta}^2}\,{\phi^2} + 
     9\,{\overline{\alpha}^2}\,{\overline{\beta}^2}\,{\phi^4} \right)
   \,{x^4}\ \    + \ \   {\overline{\alpha}^2} \,\left( 1 - {\phi^2}
   \right) \,{x^5} = 0 \ . 
 \end{eqnarray}
However not every solution of this equation corresponds to a physical
solution of a black hole. This is because the radius of the event
horizon of the Reissner-Nordstr\"om-anti-de~Sitter must obey the 
following condition
\begin{equation}
  \label{horizon}
  r_+^2-e^2+3 \alpha^2 r_+^4 \geq 0\ .
\end{equation}
Where the equality defines the extremal
Reissner-Nordstr\"om-anti-de~Sitter black hole.

Comparing (\ref{horizon}) with (\ref{betaRNADS}), yields that $\beta$ is
real and positive. Comparing it with (\ref{phiRNADS}) we obtain the
following condition
\begin{equation}
  \label{condphiRNADS}
  \phi^2 \leq \frac{1+3 \alpha^2 r_+^2}{1+\alpha^2 (1+2 r_+\, r_B + 3
    r_+^2)}\ .
\end{equation}
In the coordinates given in (\ref{newRNADS}), the inequality
(\ref{condphiRNADS}) becomes 
\begin{equation}
  \label{phi<}
  \phi^2 \leq \frac{1+3 \overline \alpha^2 x^2}{1 +\overline\alpha^2 (1+2
    x + 3 x^2)} \ .
\end{equation}
This is the condition that the solutions of equation (\ref{x^5}) must
obey in order to represent physical black hole solutions. 

Equation (\ref{x^5}) has no known analytical solutions. However its
solutions can be numerically computed and presented in graphics. 
This will be done in the next section.

\section{Analysis of the black hole solutions}
\label{s:sol}

\begin{figure}
  \begin{center}
\psfrag{x}{${}_x$}
\psfrag{b}{$\overline\beta$}
\psfrag{x}{${}_x$}
\psfrag{f}{${}_{\phi}$}
\psfrag{b  = 0.1  }{${}_{\overline\beta = 0.1}$}
\psfrag{b  = 0.3}{${}_{\overline\beta = 0.3}$}
\psfrag{b  = 0.6}{${}_{\overline\beta = 0.6}$}
\psfrag{b  = 0.9}{${}_{\overline\beta = 0.9}$}
\psfrag{b  = 3}{${}_{\overline\beta = 3}$}
\psfrag{b  = 9}{${}_{\overline\beta = 9}$}
\leavevmode
\epsfxsize=6cm
\epsfysize=4.5cm
   \centerline{\epsfbox{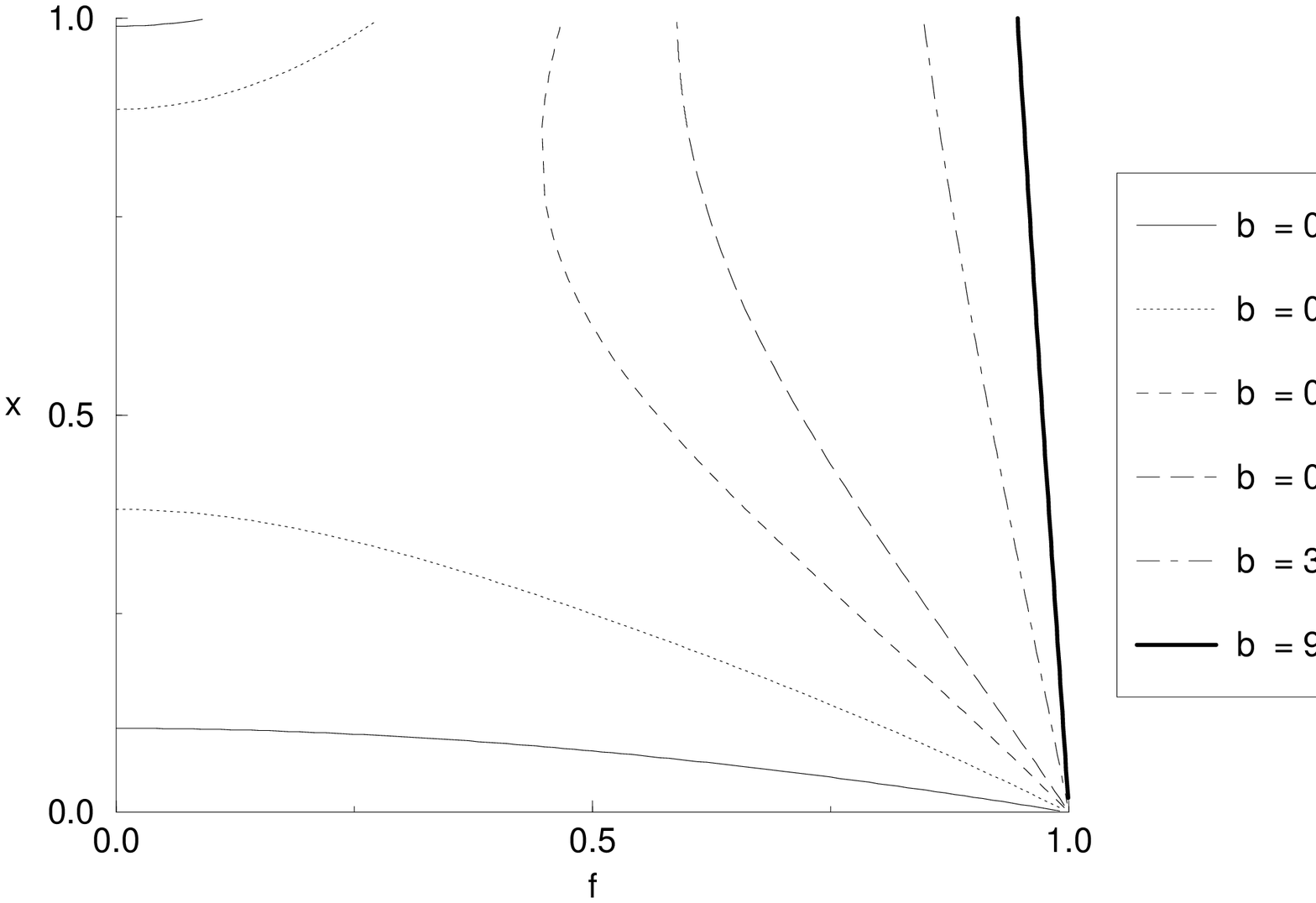}}
    \caption{Solutions of equation (\ref{x^5}) for
      $\overline \alpha=0$ (Reissner-Nordstr\"om) as a function of the
      electrostatic potential at the boundary $\phi$ for fixed values
      of $\overline \beta=0.1,0.3,0.6,0.9,3,9$. The stable solutions
      correspond to the upper branch of the curves. This means that
      when there are 2 solutions for given values of $\overline \beta$ 
      and $\phi$, only the solution with higher value of $x$ is
      stable. } 
    \label{fig1}
  \end{center}
\end{figure}

\begin{figure}
  \begin{center}
\psfrag{x}{${}_x$}
\psfrag{f}{${}_{\phi}$}
\psfrag{b  = 0.1  }{${}_{\overline\beta = 0.1}$}
\psfrag{b  = 0.3}{${}_{\overline\beta = 0.3}$}
\psfrag{b  = 0.6}{${}_{\overline\beta = 0.6}$}
\psfrag{b  = 0.9}{${}_{\overline\beta = 0.9}$}
\psfrag{b  = 3}{${}_{\overline\beta = 3}$}
\psfrag{b  = 9}{${}_{\overline\beta = 9}$}
 \leavevmode
\epsfxsize=6cm
\epsfysize=4.5cm
   \centerline{\epsfbox{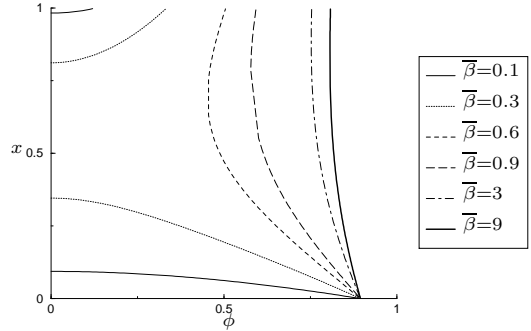}}
    \caption{Solutions of equation (\ref{x^5}) for
      $\overline \alpha=0.5$ as a function of the electrostatic
      potential at the boundary $\phi$ for fixed values of  $\overline
      \beta=0.1,0.3,0.6,0.9,3,9$. Notice that $\phi \lesssim .89$, as
      imposed by condition (\ref{phi<}). The stable solutions
      correspond to the upper branch of the curves. This means that
      when there are 2 solutions for given values of $\overline \beta$ 
      and $\phi$, only the solution with higher value of $x$ is
      stable.} 
    \label{fig2}
  \end{center}
\end{figure}

\begin{figure}
  \begin{center}
\psfrag{x}{${}_x$}
\psfrag{f}{${}_{\phi}$}
\psfrag{b  = 0.1  }{${}_{\overline\beta = 0.1}$}
\psfrag{b  = 0.3}{${}_{\overline\beta = 0.3}$}
\psfrag{b  = 0.6}{${}_{\overline\beta = 0.6}$}
\psfrag{b  = 0.9}{${}_{\overline\beta = 0.9}$}
\psfrag{b  = 3}{${}_{\overline\beta = 3}$}
\psfrag{b  = 9}{${}_{\overline\beta = 9}$}
\leavevmode
\epsfxsize=6cm
\epsfysize=4.5cm
   \centerline{\epsfbox{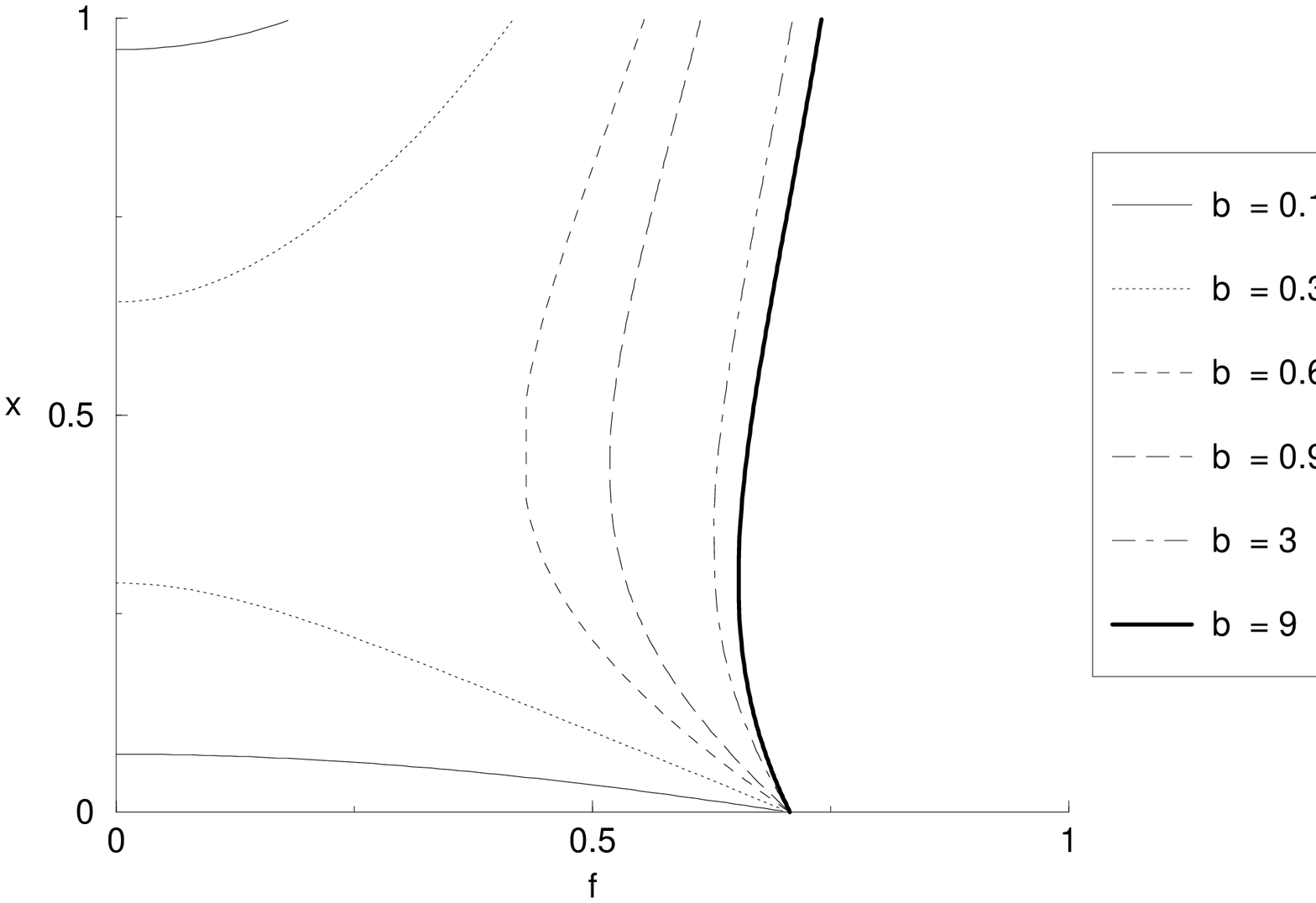}}
    \caption{Solutions of equation (\ref{x^5}) for
      $\overline \alpha=1$ as a function of the electrostatic
      potential at the boundary $\phi$ for fixed values of $\overline
      \beta=0.1,0.3,0.6,0.9,3,9$. The maximum value of $\phi$ for which
      there are solutions, $\phi_{\rm max} \simeq 0.76$ is imposed by
      condition  (\ref{phi<}). The stable solutions
      correspond to the upper branch of the curves. This means that
      when there are 2 solutions for given values of $\overline \beta$ 
      and $\phi$, only the solution with higher value of $x$ is
      stable.}  
    \label{fig3}
  \end{center}
\end{figure}

\begin{figure}
  \begin{center}
\psfrag{x}{${}_x$}
\psfrag{f}{${}_{\phi}$}
\psfrag{b  = 0.1  }{${}_{\overline\beta = 0.1}$}
\psfrag{b  = 0.3}{${}_{\overline\beta = 0.3}$}
\psfrag{b  = 0.6}{${}_{\overline\beta = 0.6}$}
\psfrag{b  = 0.9}{${}_{\overline\beta = 0.9}$}
\psfrag{b  = 3}{${}_{\overline\beta = 3}$}
\psfrag{b  = 9}{${}_{\overline\beta = 9}$}
\leavevmode
\epsfxsize=6cm
\epsfysize=4.5cm
   \centerline{\epsfbox{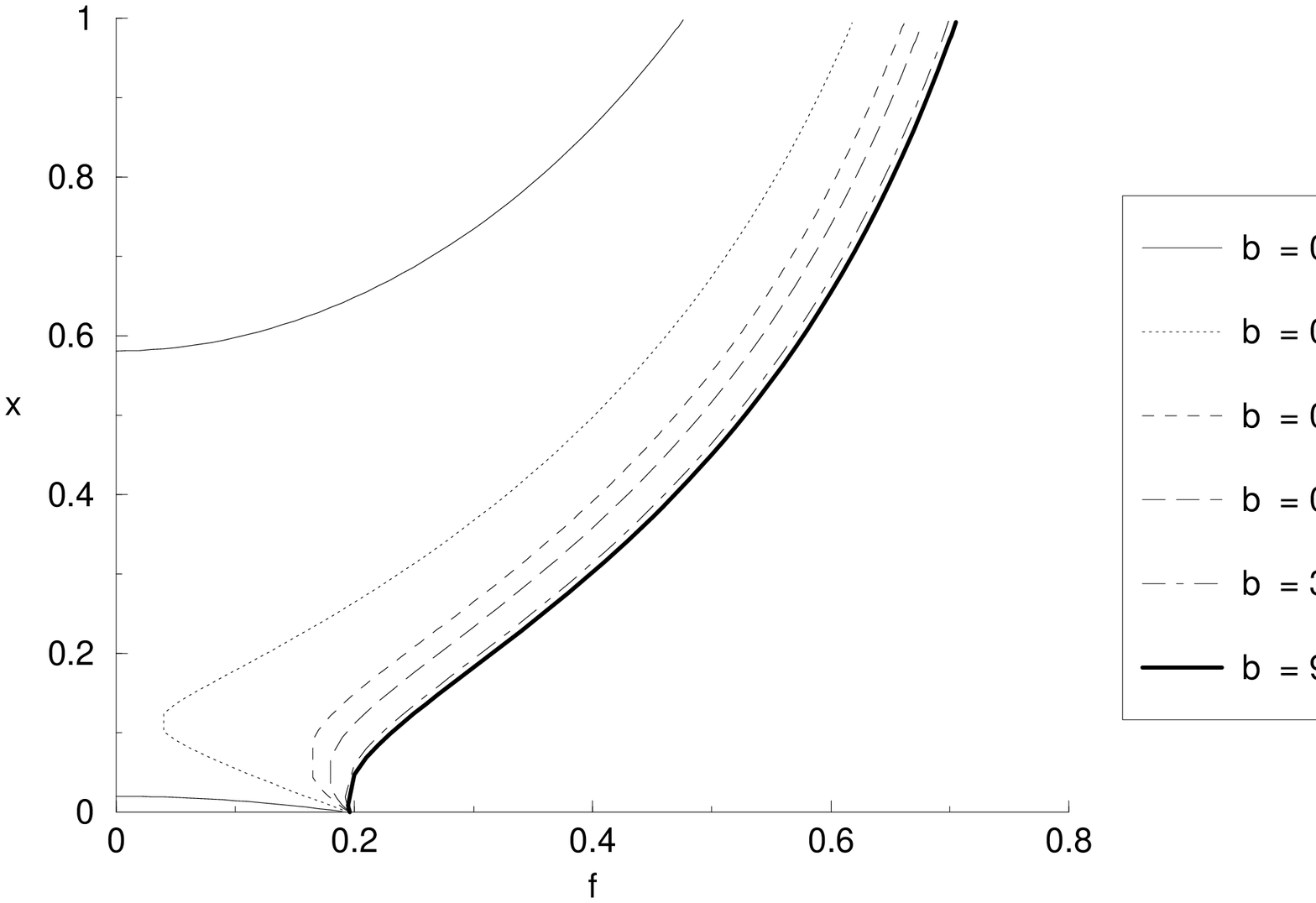}}
    \caption{Solutions of equation (\ref{x^5}) for
      $\overline \alpha=5$ as a function of the electrostatic
      potential at the boundary $\phi$ for fixed values of $\overline
      \beta=0.1,0.3,0.6,0.9,3,9$. The maximum value of $\phi$ for
      which there are solutions, $\phi_{\rm max} \simeq 0.71$, is
      imposed by condition (\ref{phi<}). The stable solutions
      correspond to the upper branch of the curves. This means that
      when there are 2 solutions for given values of $\overline \beta$ 
      and $\phi$, only the solution with higher value of $x$ is
      stable.}
    \label{fig4}
  \end{center}
\end{figure}

\begin{figure}
  \begin{center}   
   \psfrag{f}{$\phi$}
   \psfrag{a}{$a$}
\leavevmode
\epsfxsize=6cm
\epsfysize=6cm
   \centerline{\epsfbox{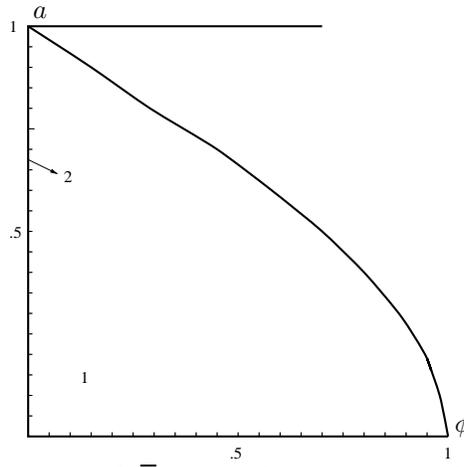}}
    \caption{Number of solutions of equation
      (\ref{x^5}) with $\overline \beta \to 0$, in the space
      $\phi \times a$, where $a$ is given by equation
      (\ref{aRNADS}). There is one black hole solution in the confined
      region and also for $a=1$ (i.e. infinite cosmological constant),
      for $\phi < \sqrt{0.5}$. There are two solutions for $\phi=0$,
      i.e. for the Schwarzschild-anti-de~Sitter black hole.}
    \label{fig5}
  \end{center}
\end{figure}

\begin{figure}
  \begin{center}
     \psfrag{f}{$\phi$}
   \psfrag{a}{$a$}
\leavevmode
\epsfxsize=6cm
\epsfysize=6cm
   \centerline{\epsfbox{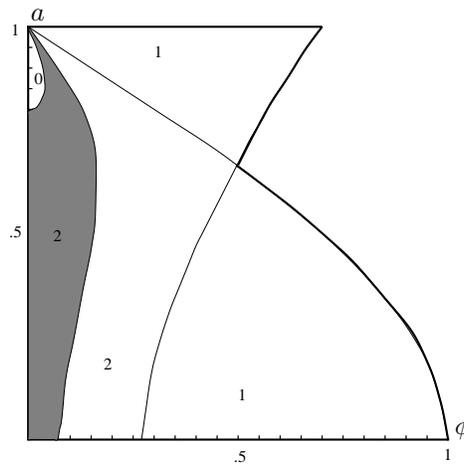}}
    \caption{Number of solutions of equation
      (\ref{x^5}) with $\overline \beta = 0.3$, in the space
      $\phi \times a$, where $a$ is given by equation
      (\ref{aRNADS})   0 - means that there are zero solutions
      in this region, i.e.  there are no solutions of black holes  in
      thermodynamical equilibrium for this set of values of $\overline
      \beta$, $\phi$ and  $a$.  1 - means that there is one solution. 2 - 
      means that there are two solutions. In the shaded
      region the stable solutions are not globally stable and
      therefore represent metastable black holes (see subsection
      \ref{global}).}
    \label{fig6}
  \end{center}
\end{figure}

\begin{figure}
  \begin{center}  
   \psfrag{f}{$\phi$}
   \psfrag{a}{$a$}
\leavevmode
\epsfxsize=6cm
\epsfysize=6cm
   \centerline{\epsfbox{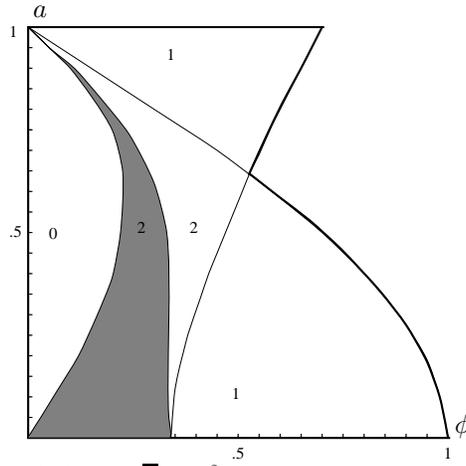}}
    \caption{Number of solutions of equation
      (\ref{x^5}) with 
      $\overline \beta = \frac{2}{\sqrt{27}} \simeq 0.38$, in the space
      $\phi \times a$, where $a$ is given by equation
      (\ref{aRNADS}) (see caption of figure \ref{fig6}).} 
    \label{fig7}
  \end{center}
\end{figure}

\begin{figure}
  \begin{center}  
   \psfrag{f}{$\phi$}
   \psfrag{a}{$a$}
\leavevmode
\epsfxsize=6cm
\epsfysize=6cm
   \centerline{\epsfbox{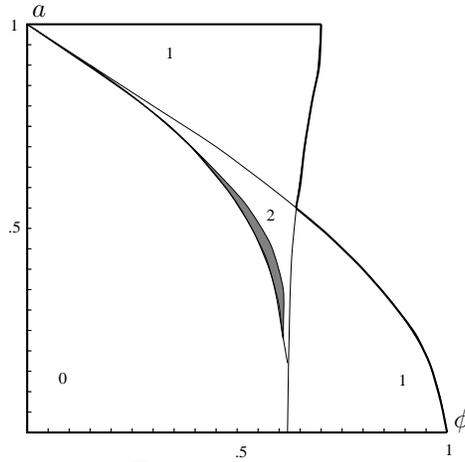}}
    \caption{Number of solutions of equation
      (\ref{x^5}) with
      $\overline \beta = 1$, in the space
      $\phi \times a$, where $a$ is given by equation
      (\ref{aRNADS}) (see caption of figure \ref{fig6}).} 
    \label{fig8}
  \end{center}
\end{figure}

\begin{figure}
  \begin{center}  
   \psfrag{f}{$\phi$}
   \psfrag{a}{$a$}
\leavevmode
\epsfxsize=6cm
\epsfysize=6cm
   \centerline{\epsfbox{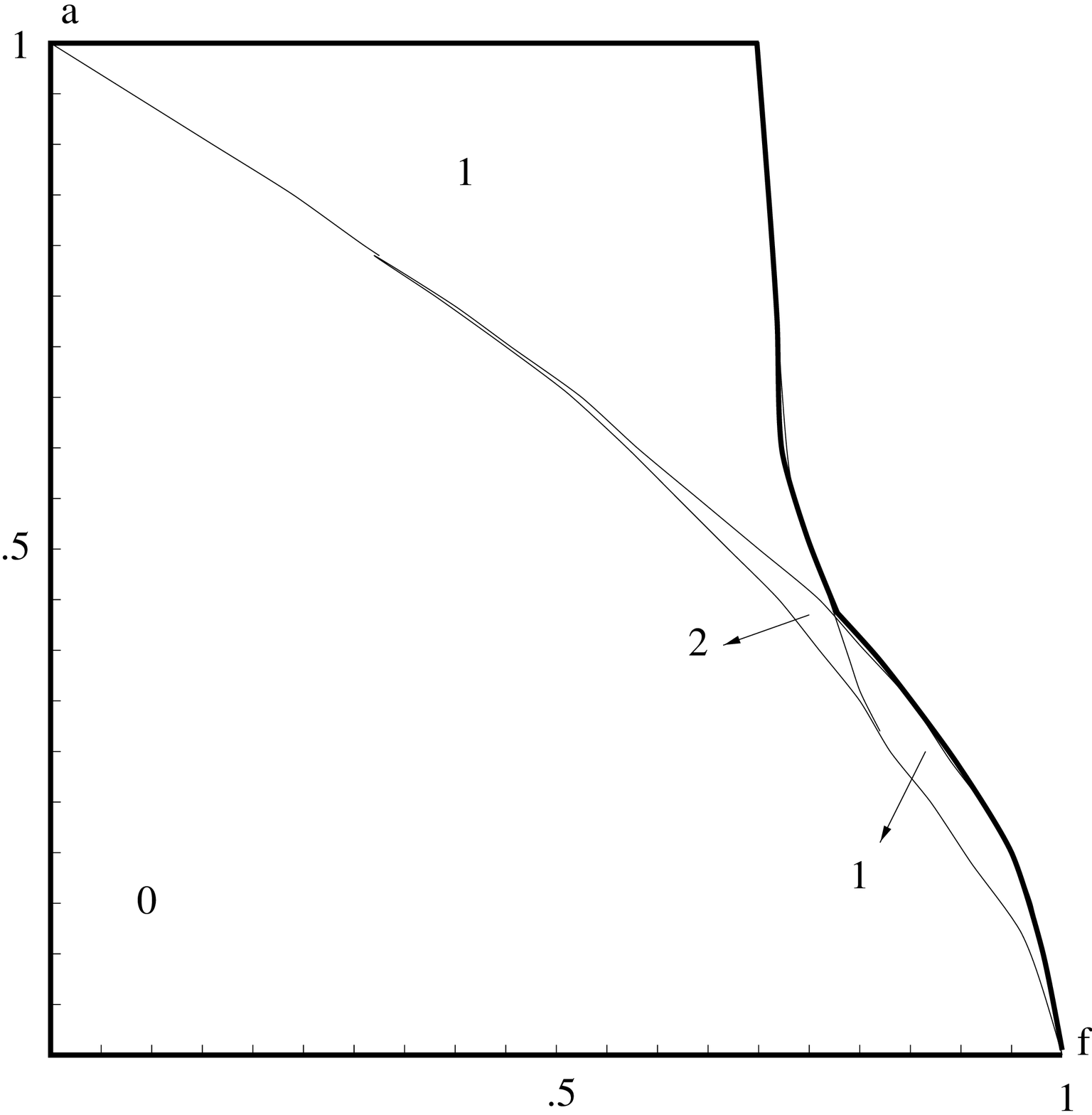}}
    \caption{Number of solutions of equation
      (\ref{x^5}) with 
      $\overline \beta = \infty$, in the space
      $\phi \times a$, where $a$ is given by equation
      (\ref{aRNADS}). Notice that for $\overline \beta\to \infty$,
      equation (\ref{x^5}) becomes (\ref{t=0}) which corresponds to
      the extremal Reissner-Nordstr\"om-anti-de~Sitter black hole, see
      discussion following equation (\ref{t=0}) (see
      caption of figure \ref{fig6}).} 
    \label{fig9}
  \end{center}
\end{figure}

In this section we present in graphics an analysis of the solutions of
equation (\ref{x^5}) that obey condition (\ref{phi<}). This analysis
is done in two steps: ($i$) first, we analyze figures
\ref{fig1} to \ref{fig4}, that present the
solutions $x$ 
as  functions of $\overline \beta$ and $\phi$, for values $\overline
\alpha=$ 0, 0.5, 1 and 5; ($ii$)
afterwards we show in figures \ref{fig5} to
\ref{fig9} the regions with  zero, one and two solutions 
in the space spanned by $\phi \times \overline \alpha$ for fixed
values of $\overline \beta$.
\vspace{.5cm}

($i$) Analysis of figures \ref{fig1} to
\ref{fig4}:

{\bf $\boldsymbol{\overline \alpha=0}$:} The solutions for $\overline
\alpha=0$ are presented in figure 
\ref{fig1}. These are obviously identical to the solutions
of the Reissner-Nordstr\"om black hole, see \cite{Braden}. We can see
in figure \ref{fig1} 
that for fixed $\phi$ there is a maximum of $\overline \beta$,
$\overline \beta_{\rm max}(\phi)$,  so that for  $\overline
\beta>\overline \beta_{\rm max}(\phi)$ there are no solutions. For
$\overline \beta<\overline \beta_{\rm max}(\phi)$  one can have two
or only one solution depending on the precise value of $\overline
\beta$. For $\phi=0$ (Schwarzschild) one has always two solutions for
$\overline \beta<\overline \beta_{\rm max}(0)$ ($\overline \beta_{\rm
  max}(0)=2/\sqrt{27}$). In the limiting case $\overline \beta \to 0$
(i.e. $r_B\,T \to\infty$), one finds there is a solution with
$x=r_+/r_B \to 1$ (as will be seen in section \ref{stab} this is the
stable solution), see \cite{York}. 
For $\phi=0$, still, and $\overline \beta>\overline \beta_{\rm
  max}(0)$ there are no solutions (see comments at the end of this
section). For $0<\phi<1/\sqrt{3}$ one has one or two solutions up to
$\overline \beta_{\rm max}(\phi)$, whereas for $\overline
\beta>\overline \beta_{\rm max}(\phi)$ there are as well no
solutions. Finally for $\phi>1/\sqrt{3}$ there is only one solution
(again for $\overline \beta<\overline \beta_{\rm max}(\phi)$)
corresponding to the unstable branch as will be seen in section \ref{stab}. 
Note that, for $\overline \alpha=0$, condition (\ref{phi<}) implies that
the electrostatic potential has a maximum at $\phi_{\rm
  max}=1$. Notice also that in the limit $\overline \beta \to \infty$
($T \to 0$), the curves in figure \ref{fig1} tend to the curve
$\phi=1$, which corresponds to the extremal Reissner-Nordstr\"om black
hole ($r_+=e$). 

The cases $\overline \alpha \neq 0$, which are now going to be analyzed
require the following prior analysis. 

As in the case $\overline \alpha=0$, for $\overline \alpha \neq 0$
there are solutions at $T=0$ ($\overline \beta = \infty$), that
correspond to the extremal black holes. This can be
analytically verified by replacing $\overline \beta= \infty$ in
equation (\ref{x^5}), from where we obtain the equation
\begin{equation}
  \label{t=0}
  1 - \phi^2 - \overline \alpha^2 \phi^2 - 2 \overline \alpha^2
  \phi^2 x +3 \overline \alpha^2 (1-\phi^2) x^2 = 0 \ .
\end{equation}
Notice this is the equation one obtains taking the equality  in
condition (\ref{phi<}). In fact it corresponds to the 
condition of extremality of the Reissner-Nordstr\"om-anti-de~Sitter
black hole, which is in agreement with the well known fact that
only the extremal black holes have zero temperature. 

Equation (\ref{t=0}) has at least
one solution that verifies $0<x<1$ if 
\begin{eqnarray}
  \label{int1RNADS}
  \overline \alpha^2 <\frac{1}{6} &\ \ \  {\rm and}\ \ \  &
  \frac{1+3\,\overline \alpha^2}{1+6\, \overline \alpha^2} \leq \phi^2 \leq
  \frac{1}{1+\overline \alpha^2}\ , \nonumber \\ \frac{1}{6} \leq \overline
  \alpha^2 < \frac{2}{3} & \ \ \ {\rm and}\ \ \  & \frac{3\, \overline
    \alpha^2 +6 - \sqrt{9\, \overline \alpha^4 +12 \, \overline
      \alpha^2}}{4\, \overline \alpha^2+6} \leq \phi^2 \leq
  \frac{1}{1+\overline \alpha^2}\ , \nonumber \\ \overline \alpha^2 \geq
  \frac{2}{3} & \ \ \ {\rm and}\ \ \  & \frac{3\, \overline
    \alpha^2 +6 - \sqrt{9\, \overline \alpha^4 +12 \, \overline
      \alpha^2}}{4\, \overline \alpha^2+6} \leq \phi^2 \leq
  \frac{1+3\,\overline \alpha^2}{1+6\, \overline \alpha^2} \ .
\end{eqnarray}
For these values of $\phi$ and $\overline \alpha$, the curves for
fixed $\phi$ presented in the figures reach infinite $\overline
\beta$. 
Furthermore, equation (\ref{t=0}) has two solutions if
\begin{eqnarray}
  \label{int2RNADS}
1/6<\overline \alpha^2<2/3 &\ \ \  {\rm and}\ \ \  &
  \frac{3 \overline \alpha^2 +6 - \sqrt{9 \overline \alpha^4 + 12
      \overline \alpha^2}}{4 \overline \alpha^2+6} < \phi^2 <
  \frac{1+3 \overline \alpha^2}{1+6 \overline \alpha^2} \ , \nonumber \\
\overline \alpha^2 \geq 2/3 & \ \ \ {\rm and} \ \ \ & 
  \frac{3 \overline \alpha^2 +6 - \sqrt{9 \overline \alpha^4 + 12
      \overline \alpha^2}}{4 \overline \alpha^2+6} < \phi^2
  < \frac{1}{1+\overline \alpha^2} \ .
\end{eqnarray}

{\bf $\boldsymbol{\overline \alpha=0.5}$:} The solutions for $\overline
\alpha=0.5$ are presented in figure \ref{fig2}. 
This figure presents the same  properties mentioned previously for the case
$\overline \alpha=0$. Comparing \ref{fig1} and
\ref{fig2}, we verify that the maximum value of
$\overline \beta$, $\overline \beta_{\rm max}(\phi)$, for which there
are solutions is increasing, i.e. there are solutions for slightly
lower values of the temperature at $\overline \alpha=0.5$ than at
$\overline \alpha=0$.
 Furthermore there are solution for infinite
$\overline \beta$ in the interval $0.83 \lesssim \phi \lesssim 0.89$,
see (\ref{int1RNADS}), this can be seen using the curve $\overline
\beta =9$ in figure \ref{fig2}, since it corresponds to a good
approximation of infinite $\overline \beta$.
 
Condition (\ref{phi<}) implies, for $\overline \alpha^2 < 2/3$,
\begin{equation}
  \label{phi_0}
  \phi \leq \sqrt{\frac{1}{1+\overline \alpha^2}} \ .
\end{equation}
This condition corresponds to the upper-limit of the interval given in
(\ref{int1RNADS}), which in this case is $\phi \simeq 0.89$. 
Notice that if in equation (\ref{x^5}) we do $x=0$, we obtain
precisely $\phi=1/\sqrt{1+\overline \alpha^2}$, as can be seen in
figure \ref{fig2}.

{\bf $\boldsymbol{\overline \alpha=1}$:}  Figure
\ref{fig3} plots the solutions for $\overline
\alpha=1$. We can see that this figure presents similar
properties to the ones obtained for lower values of $\overline
\alpha$. Using (\ref{int1RNADS}), we can see that for $\phi \lesssim
0.66$ there are solutions only for $\overline \beta < \overline
\beta_{\rm max}(\phi)$, where  $\overline \beta_{\rm max}(\phi)$ is 
finite and depends on $\phi$. On the contrary, the curves
for higher values of $\phi$ reach infinite $\overline \beta$. In
particular, for $0.66 \lesssim \phi \lesssim 0.71$ (see equation
(\ref{int2RNADS})) there are two solutions at low temperatures (i.e.,
high $\overline \beta$) as can be seen in figure
\ref{fig3}, since the curve $\overline \beta=9$ is
representative of the curves with high $\overline \beta$. It can be
seen that for $\overline \beta=9$ there are two solutions for $\phi <
\phi_0$, where $\phi_0= \frac{1}{\sqrt{1+\overline\alpha^2}} \simeq
0.71$ is the value of $\phi$ where $x=0$ for every $\overline
\beta$. There is one solution for $0.71 \lesssim \phi \lesssim 0.76$,
see figure \ref{fig3}, where the upper-limit is imposed by
condition (\ref{phi<}). In fact, condition (\ref{phi<}) imposes, for
$\overline \alpha^2 <2/3$, 
\begin{equation}
  \label{phimax}
  \phi_{\rm max} = \sqrt{\frac{1+3\, \overline \alpha^2}{1+6\,
      \overline \alpha^2}} \ .
\end{equation}
Notice this is the upper-limit of the interval given in
(\ref{int1RNADS}).

{\bf $\boldsymbol{\overline \alpha=5}$:} Figure 
\ref{fig4} presents the solutions for $\overline
\alpha=5$. Using (\ref{int1RNADS}), we can see that for $\phi \lesssim
0.19$, there are solutions only for $\overline \beta < \overline
\beta_{\rm max}(\phi)$. For $0.19 \lesssim \phi \lesssim 0.71$ there is
one solution for high values of $\overline \beta$ (consider the curve
$\overline \beta=9$).  For infinite $\overline \beta$ this region is $0.195
\lesssim \phi \lesssim 0.196$, see (\ref{int2RNADS}). In figure
\ref{fig4}, we can also see that for $0.19 \lesssim \phi
\lesssim 0.71$ there is one solution, where the upper-limit is given
by (\ref{phimax}).

For higher values of $\overline \alpha$ there are not new types of
solutions and therefore it is not necessary to pursue our
analysis. 

\vspace{.5cm} 
($ii$) Analysis of figures \ref{fig5}
to \ref{fig9}:

In order to clarify the disposition of the number of solutions for
given values of $\overline \alpha$, $\phi$, and $\overline \beta$, we
present, in the space spanned by $\overline \alpha$ and $\phi$ for fixed
$\overline \beta$, the regions with zero, one and two solutions. We do
this for eleven different values of $\overline \beta$, $\overline
\beta=0, 0.3, 2/\sqrt{27}\simeq 0.38, 1, \infty$, see figures
\ref{fig5} to \ref{fig9} respectively. 
In this figures  one can see the evolution of the number of solutions
as $\overline \beta$ increases.  To present all possible values of
$\overline \alpha \epsilon [0,\infty[$, we use in these figures the
parameter $a$ instead of $\overline\alpha$, $a$ is defined by
\begin{equation}
  \label{aRNADS}
  a= \frac{2}{\pi}  \arctan{\overline \alpha} \ ,
\end{equation}
such that $0\leq a \leq 1$. 
It is this variable that
appears in the ordinate axis in figures \ref{fig5} to
\ref{fig9}.

Due to condition (\ref{phi<})  there are no physically possible
solutions on the right-hand side of figures \ref{fig5} to
\ref{fig9}. 

An important value of $\overline \beta$, first studied by York
\cite{York} in connection with the Schwarzschild black hole ($\phi=\overline
\alpha=0$),  is $\overline \beta=2/\sqrt{27}\simeq 0.38$,
i.e. $\beta= \frac{8 \pi}{\sqrt{27}}\,r_B$. For higher values of
$\beta$, lower values of the temperature, there are no black hole
solutions. This is a quantum effect and following York \cite{York}
can be understood as follows. One can associate a Compton type
wavelength $\lambda$ to the energy $k_B T$ of the thermal particles,
by $\lambda= \frac{\hbar c}{k_B T}$, or in Planck units
$\lambda=1/T=\beta$. If this Compton wavelength is much larger than the
radius $r_B$ of the cavity (or more specifically $\lambda >\frac{8
  \pi}{\sqrt{27}}\,r_B \simeq 4.8\, r_B$) then the thermal particles
cannot be confined within the cavity and do not collapse to form a
black hole. 

By analyzing figures \ref{fig6} and \ref{fig7}, we can
see that for nonzero cosmological constant  ($\overline \alpha \neq
0$) this phenomenon starts at even lower $\overline \beta$ (higher
$T$). Indeed using equation (\ref{betarnads}) (with $\phi=0$,
i.e. $q=0$) we can show that to first order in $\alpha^2$ ($\alpha^2
\, r_B^2 <<1$), York's criterion for no black hole solutions becomes 
\begin{equation}
  \label{lamb>}
  \lambda = \beta > \frac{8 \pi}{\sqrt{27}}\, r_B \left(1-\frac{5}{18}
    \alpha^2 \, r_B^2 \right) \ .
\end{equation}
From equation (\ref{lamb>}) we infer that the role of the negative
cosmological constant ($\Lambda=-3 \alpha^2$) is to produce an
effective cavity radius $r_{\rm eff} = r_B \left(1-\frac{5}{18}
  \alpha^2 \, r_B^2 \right)$ smaller than $r_B$. Thus 
for a given temperature, it is more difficult to confine the thermal
particles, and harder to form black holes, in accord with the idea
that a negative cosmological constant shrinks space. 

If we extend our previous first order analysis to 
include the electrostatic potential $\phi$, we obtain 
\begin{equation}
  \label{lam>}
    \lambda = \beta > \frac{8 \pi}{\sqrt{27}}\, r_B \left(1-\frac{5}{18}
    \alpha^2 \, r_B^2 + 2 \phi^2 \right) \ .
\end{equation}
We can see that the electrostatic potential has the opposite
effect of the cosmological constant (see for example figures
\ref{fig6}, \ref{fig7} and \ref{fig8}).

\section{Stability}
\label{stab}

\subsection{Local stability}
As mentioned before there is a second condition of local stability that
has not yet been investigated. We will follow the same procedure as
given in \cite{Braden}. This is the condition that the Hessian 
matrix of the reduced action  be positive definite. 
For convenience in this analysis we will use the variable $S=\pi
r_+^2$ instead $r_+$. The Hessian matrix of $I^*(S,e)$ is 
\begin{equation}
\label{I,ij}
I^*_{,ij}=\left( \begin{array}{cc} I^*_{,ee}&I^*_{,eS} \\  I^*_{,eS}&I^*_{,SS}
\end{array} \right)\ .
\end{equation}
The matrix is positive definite if its pivots are all positive. The
pivots of (\ref{I,ij}) are  
\begin{equation}
\label{pivot1}
I^*_{,ee}
\end{equation}
and
\begin{equation}
\label{pivot2}
\frac{{\rm det}(I^*_{,ij})}{I^*_{,ee}}  \ .
\end{equation}
The first condition of local stability, $\nabla I^*=0$ yields
\begin{eqnarray}
\label{term}
\left(\frac{\partial I^*}{\partial e}\right)_S&=&\beta
\left(\frac{\partial E^*}{\partial e}\right)_S-\beta \, \phi=0 \ \ \ 
\Rightarrow \ \ \ \phi=\left(\frac{\partial E}{\partial e}\right)_S \ ,
\nonumber \\ \left(\frac{\partial I^*}{\partial S}\right)_e &=&\beta
\left(\frac{\partial E^*}{\partial S}\right)_e -1=0 \ \ \
\Rightarrow \ \ \ \beta^{-1}=T= \left(\frac{\partial E}{\partial
S}\right)_e \ .
\end{eqnarray}
These are well known thermodynamical relations \cite{Callen}. 

From (\ref{term}), we compute the  second derivatives of $I^*$ in
the stationary points of $I^*$
\begin{eqnarray}
\label{2der}
\left.\frac{\partial^2 I^*}{\partial e^2}\right|_{\rm eq}&=&\left.\beta\,
\frac{\partial^2 E^*}{\partial e^2}\right|_{\rm eq}=
\beta \left(\frac{\partial \phi}{\partial e}\right)_S \nonumber \\
\left.\frac{\partial^2 I^*}{\partial e\,\partial S}\right|_{\rm eq}&=&
\left.\beta \, 
\frac{\partial^2 E^*}{\partial e\,\partial S}\right|_{\rm eq}=
\beta \left(\frac{\partial \phi}{\partial S}\right)_e =
-\frac{1}{\beta} \,\left(\frac{\partial \beta}{\partial e}\right)_S
\nonumber \\ \left.\frac{\partial^2 I^*}{\partial S^2}\right|_{\rm eq}&=&
\left. \beta\,
\frac{\partial^2 E^*}{\partial S^2} \right|_{\rm eq}=
-\frac{1}{\beta} \left(\frac{\partial \beta}{\partial S}\right)_e \ .
\end{eqnarray}
where ${\rm eq}$ means quantities evaluated at equilibrium, i.e. at the
stationary points of the reduced action $I^*$. 

The first pivot (\ref{pivot1}) is simply the first of these
equations. The second pivot (\ref{pivot2}) is easily obtained from
(\ref{2der}) 
\begin{eqnarray}
\label{pivot3}
\frac{{\rm det} I^*_{,ij}}{I^*_{,ee}}&=&-\frac{1}{\beta}\, \left(
  \frac{\partial \beta}{\partial S}\right)_e+\frac{1}{\beta} \,
\left(\frac{\partial\beta}{\partial e}\right)_S\, \left(\frac{\partial
    \phi}{\partial S}\right)_e\,\left/ \left(\frac{\partial
      \phi}{\partial e}\right)_S \right. \nonumber \\ &=&
-\frac{1}{\beta}\, \left[ \left(\frac{\partial \beta}{\partial S}
\right)_e  + \left(\frac{\partial \beta}{\partial e}\right)_S \,
\left(\frac{\partial e}{\partial S}\right)_\phi \right] \nonumber \\ &=&
-\frac{1}{\beta} \left(\frac{\partial \beta}{\partial S}\right)_\phi =
\frac{1}{C_{\phi,r_B}}\ ,
\end{eqnarray}
where $C_{\phi,r_B}$ is the heat capacity at constant $\phi$ and
$r_B$.
Notice that in all this calculation we have implicitly held $r_B$
constant, since in the grand canonical ensemble the dimension of the
system is held constant. 

Imposing positive pivots for  local stability  yields
\begin{eqnarray}
\label{est}
   \left(\frac{\partial \phi}{\partial e}\right)_{S,r_B} \geq 0  \ ,
   \nonumber \\
 C_{\phi,r_B} \geq 0 \ .
\end{eqnarray}
These conditions are identical to the classical thermodynamical
stability conditions \cite{Callen}. Therefore one can conclude
 that the dynamical  stability conditions given in
(\ref{term}) and (\ref{est}) are identical to the thermodynamical
stability conditions \cite{Braden}.  

For the Reissner-Nordstr\"om-anti-de~Sitter black hole, the pivots
obtained from (\ref{2der}) and (\ref{pivot3}) are
\begin{eqnarray}
  \label{est1RNADS}
\left( \frac{\partial \phi}{\partial e} \right)_{S,r_B} &=&
  \left(\frac{1}{r_+}-\frac{1}{r_B} \right) \left(
    1-\frac{r_+}{r_B}-\frac{e^2}{r_+ r_B}-\frac{\alpha^2
      {r_+}^3}{r_B}+ \frac{e^2}{{r_B}^2}+\alpha^2 {r_B}^2
  \right)^{1/2}+ {} \nonumber \\
 &{}& +\frac{e^2}{r_B} \left(\frac{1}{r_+}-\frac{1}{r_B}
  \right)^2 \left(1-\frac{r_+}{r_B}-\frac{e^2}{r_+ r_B}-\frac{\alpha^2
      {r_+}^3}{r_B}+ \frac{e^2}{{r_B}^2}+\alpha^2 {r_B}^2
  \right)^{3/2} \ , \nonumber \\ 
\end{eqnarray}
which is positive,  and
\begin{eqnarray}
  \label{est2RNADS}
C_{\phi,r_B} &=& 4 \pi {r_+}^3 (r-r_+)({r_+}^2-e^2+3 \alpha^2
  {r_+}^4)[1+\alpha^2 ({r_B}^2+r_+ r_B+{r_+}^2)] \times {}
\nonumber \\ {} &\times& \left\{e^4+{r_+}^3 [(6
  \alpha^2 {r_+}^2-2) (r_B+\alpha^2 {r_B}^3)+3 r_++2 \alpha^2
  {r_+}^3+3 \alpha^4 {r_+}^5]+ \right.{} \nonumber \\ {} &+& \left. 2
  e^2 r_+ [-2 r_+ 
  + r_B (1+\alpha^2 ({r_B}^2-2 r_+ r_B -2 {r_+}^2))]\right\}^{-1} \ .
\end{eqnarray}
The numerator of $C_{\phi,r_B}$ is positive, therefore the condition
$C_{\phi,r_B}>0$ is verified if the denominator in (\ref{est2RNADS}) is
positive. Using (\ref{newRNADS}) and (\ref{q(x)RNADS}), we obtain the
following condition of stability for the solutions of equation
(\ref{x^5})
\begin{eqnarray}
  \label{estRNADS}
  &&\left[ -2 \,(1+\overline \alpha^2)+3\, x+6 \,\overline \alpha^2
    (1+\overline \alpha^2) \,x^2 +2 \,\overline \alpha^2 x^3 + 3 \,\overline
    \alpha^4 x^5 \right](1-x+x \phi^2)^2 -{} \nonumber \\ {}&-&\left[ 2\,
    \phi^2 (1+\overline \alpha^2 (1+x+x^2))(-1 + 2\, x +\overline
    \alpha^2 (-1+2 x+2 x^2))\right] (1-x+x \phi^2)+{} \nonumber \\ {}
  &+& \phi^4 x \,(1+\overline \alpha^2 (1+x+x^2))^2 \ > \ 0 \ .
\end{eqnarray}
By numerical computation, we can verify the solutions that obey this
condition. We have found that the lower branches of the curves
presented in figures \ref{fig1} to \ref{fig4} correspond to unstable
solutions, while the upper branches correspond to stable solutions. 
Therefore  we can say that in general only the
solutions with the  higher value of $x$, that is with higher event
horizon radius, are stable. 

In more detail we can distinguish three
cases: ($i$) for low $\overline \alpha$ (see figures
\ref{fig1} and \ref{fig2}), and for values of $\beta$ and $\phi$ for
which there are one solution, it corresponds to a lower branch and
therefore this solutions is unstable; ($ii$) for high values of
$\overline \alpha$ (see figure \ref{fig4}), whenever there is only
one solution, it corresponds to an upper branch and therefore this
solution is stable; ($iii$) whenever there are two solutions,
for given values of $\overline \beta,\phi, \overline \alpha$, the
smaller one is an instanton (i.e. it is an unstable solution and
dominates the semi-classical evaluation 
of the rate of nucleation of black holes \cite{Yaffe}) and the one with larger
event horizon radius is a stable black hole.  

These three cases can easily be distinguished in figures
\ref{fig5} to \ref{fig9}. In these figures there are in
general two separated regions with one solution. The region with lower
values of $a$, i.e. lower values of $\overline \alpha$, corresponds to
case ($i$) and these are unstable solutions. The region with
higher values of $a$ corresponds to case ($ii$) and these are
stable solutions. Obviously, the region with two solutions in each
figure \ref{fig5} to \ref{fig9}, corresponds to case
($iii$), i.e. one of those solution is stable, the one with higher
value of $x$, and the other unstable. 

\subsection{Global stability}
\label{global}

The stable solutions computed in the previous subsection are not
necessarily global minimum of the action. In this case they will not
dominate the partition function and the zero-loop approximation cannot
be considered accurate \cite{WhitYork,Braden}. 

As the reduced action given in equation (\ref{IRNADS}) grows without
bound in the  directions where $r_+$ or $e$ tend to infinity, the
global minimum must be found either at the local minimum or at
$r_+=e=0$. The action tends to zero as $r_+$ and $e$ tend to zero,
therefore the condition of global stability of the stable
solutions is that the classical action $I$ must be negative. 

If the classical action is positive, the partition function is
dominated by points near the origin. But these points do not
correspond to a black hole in thermal equilibrium. In this case, the
black hole, that corresponds to the stable solution of the reduced
action, is metastable. 

We verified numerically which boundary conditions (given by the values
of $\beta$, $\phi$ and $\alpha$) correspond to globally stable black
holes, i.e. to solutions that dominate the partition function. 
We can also see, using a simple argument, that for $\overline \beta <
\frac{8}{27}$ all locally stable solutions are also globally stable. 
Indeed, York \cite{WhitYork} has shown that for the Schwarzschild black 
hole the condition of global stability is $\overline \beta <
\frac{8}{27}$ and since the classical action $I$ is a decreasing
function of $\phi$ and $\alpha$, this condition is still a bound for
globally stability for all $\phi$ and $\alpha$. However for
$\overline \beta > \frac{8}{27}$, there is always a certain region of
$\phi\times\alpha$, for which the locally stable solutions do not
correspond to global minima of the action. 
We show these regions in figures \ref{fig5} to \ref{fig9}, where the
regions for which the solutions do not correspond to a global minimum
of the action, i.e. the metastable solutions, are shaded. In
particular figures \ref{fig5} and \ref{fig9} do not present a shaded
region because all the stable solutions are dominant in the limit
$\overline \beta \to 0$ as said above and for $\overline \beta \to
\infty$ the region with metastable solutions is too thin to be
presented in graphic.

\section{The $r_B \to \infty$ limit and the Hawking-Page solutions}
\label{s:infty}

One can study the case where the boundary goes to infinity. There are
two different ways for taking this limit: ($i$) fixing
the horizon radius $r_+$ and the charge $e$ of the black hole; ($ii$)
fixing the boundary conditions, i.e. fixing $\beta$ and $\phi$.
In this section we will study only the pure
Reissner-Nordstr\"om-anti-de~Sitter cases, i.e. the cases with
$\alpha, \phi \neq 0$. Other cases, with either $\alpha$ or $\phi$
equal to zero are considered in the next section. 

We will start by studying the first case: (i)
Fixing the black hole solution, i.e. fixing $r_+$
and $e$ and taking the limit $r_B \to \infty$, we obtain from equations
(\ref{betaRNADS}) and (\ref{phiRNADS}) that the temperature
$T=\beta^{-1}$ and electrostatic potential $\phi$ go to zero as $T
\sim \frac{r_+^2-e^2+3\,\alpha^2\,r_+^4}{4\,\pi\,\alpha\,r_+^3} \,
r_B^{-1}$ and $\phi \sim \frac{e}{\alpha r_+} \, r_B^{-1}$,
respectively. In this case the thermal energy goes to zero as 
$  E \sim M \, \left(1+ \alpha^2 \, r_B^2 \right)^{-1/2}$,
where $M$ is the mass of the black hole given in (\ref{2M}). As also
previously found in \cite{Brown94}, the thermal energy at infinity is
not equal to the ADM mass of the black hole, due to the the fact that
the spacetime is not asymptotically flat. Note that in
\cite{Louko,HP83} it is found that $E=M$ due to a different definition
of the temperature of the ensemble. 

To determine the stability of the black hole solutions in this limit,
we compute the heat capacity, from equation (\ref{est2RNADS}) and
obtain 
\begin{equation}
  \label{Cinfty}
  C_{\phi} = \frac{2\, \pi\, r_+^2 \left( r_+^2-e^2+3\,\alpha^2\,r_+^4
      \right)}{e^2 - r_+^2 +3\,\alpha^2\,r_+^4} \ .
\end{equation}
The numerator in (\ref{Cinfty}) is necessarily positive due to condition
(\ref{horizon}), that $r_+$ must obey to represent the event horizon
radius. Therefore, the stability condition for these solutions is
simply $e^2 - r_+^2 + 3\,\alpha^2\,r_+^4 >0$. 
Notice that equation (\ref{Cinfty}) is a generalization to $e \neq 0$
of the heat capacity found by Hawking and Page in \cite{HP83}. 
Computing the action in this limit yields
\begin{equation}
  \label{Iinfty}
  I= \frac{\pi\, r_+^2 \left( r_+^2- e^2 -\alpha^2\,r_+^4
      \right)}{ r_+^2 - e^2 +3\,\alpha^2\,r_+^4} \ .
\end{equation}
This is precisely Hawking-Page action \cite{HP83,Louko}, which here 
we have recovered from York's formalism taking the appropriate limit. 

From equation (\ref{Iinfty}), we can verify the global stability of
the black hole by imposing that $I<0$. Therefore we conclude that the
solution given by ($r_+,e$) corresponds to a globally stable black
hole if conditions $r_+^2- e^2 -\alpha^2\,r_+^4<0$ and (\ref{horizon})
are both verified. These conditions are similar to those found in
\cite{Louko}.

We conclude that taking different boundary conditions, i.e. choosing to fix
the boundary conditions in the boundary at infinity (as done here) or
in the region where spacetime is flat like Hawking and Page
\cite{HP83}, yields a different value for the energy only. Furthermore this
difference is so that all the other physical quantities (like the action,
entropy, mean-value of the charge and heat capacity) remain the same.

On the other hand, we can take a different limit: 
(ii) we can fix the boundary conditions, i.e. $\beta$ and
$\phi$, when taking the limit. This is done by recovering the variables
$\beta$ and $\alpha$ in equation (\ref{x^5}), using (\ref{newRNADS}). 
Taking the limit $r_B \to \infty$, we obtain the equation
$  \alpha^2 \, \beta^2 \, \left( \phi^2 \, (1+2 \, x+3 \, x^2) -3 \,
    x^2\right)^2 -x^2 \, (1+x+x^2) \, (1-x+\phi^2 \, x) =0$. 
This is a 5th degree equation in $x$.  For fixed 
$\beta$, $\phi$ and $\alpha$ and taking the limit $r_B \to \infty$, we
obtain a solution $r_+$ that tends to infinity like $r_+ \sim c_+ r_B$,
where $c_+$ is a constant that depends only on $\beta$, $\phi$ and
$\alpha$. 
Considering now equation (\ref{q(x)RNADS}), we can see that the charge $e$
goes to infinity as $e \sim  c_e r_B^2$, where again $c_e$ is a
constant that depends only on $\beta$, $\phi$ and $\alpha$. Therefore
the entropy (\ref{SRNADS}) and the mean-value of the charge
(\ref{QRNADS}) both go to infinity as $r_B^2$. In this limit the
action, the thermal energy and the heat capacity, given in
(\ref{IRNADS}), (\ref{ERNADS}) and (\ref{est2RNADS}) respectively,
also diverge as $r_B^2$. The heat capacity is always positive, which
means these solutions are stable. The action is always positive, therefore
the solutions are not globally stable and represent metastable black
holes.

\section{Comments on special cases}
\label{s:special}

Several black holes may be considered as special cases of the
Reissner-Nordstr\"om-anti-de~Sitter black hole. 
\begin{description}

\item[(I)] putting $\phi=0$ and $\Lambda=0$, we obtain the
  Schwarzschild black hole studied in \cite{York}. There are two
  solutions for $\overline \beta<\overline \beta_{\rm
    max}=2/\sqrt{27}$.  This solutions can be computed analytically,
  since equation (\ref{x^5}) becomes a 3rd degree equation for
  $\Lambda=0$. Only the solution with higher event horizon radius,
  i.e. higher mass, is stable. 

In the limit $r_B \to \infty$ the unstable solution, i.e. the solution
with lower value of the horizon radius, goes to $r_+=1/(4\pi r_+)$,
while the stable solution goes to infinity as $r_+ \sim r_B$
\cite{York}. 

\item[(II)] putting $\Lambda=0$ we obtain the Reissner-Nordstr\"om
  black hole. This has been studied in \cite{Braden}. There are one or
  two solution for $\overline \beta<\overline \beta_{\rm max}$. These
  as stated above can be computed analytically. Once again only  the
  solution with higher event horizon radius, when it exists, is
  stable. 

In the limit $r_B \to \infty$, the black hole horizon radius is $r_+=
\frac{\beta}{4 \pi} \left(1-\phi^2\right)$ and the charge is given by
$e= \phi r_+$. The thermal energy of these solutions is equal to their
ADM mass $E=M$. The heat capacity is negative $C_\phi=-2\pi
r_+^2$. therefore the solutions are unstable. 

\item[(III)] putting $\phi=0$ we obtain the
  Schwarzschild-anti-de~Sitter. This black hole has been studied
  before in \cite{HP83,Brown94}. This black hole
  has two solutions for $\beta<\beta_{\rm max}$, and again only the
  one with higher event horizon radius is stable. 

The limit $r_B \to \infty$ can be taken in two ways: ($i$) fixing the
temperature and the cosmological constant, there are one unstable
solution that tends to zero as $r_+ \sim \frac{\beta}{4 \pi \alpha}
r_B^{-1}$ and one stable solution that tends to infinity as $r_+ \sim
c r_B$, where $c$ is the solution of equation $c^3 + \left(\frac{3
    \alpha \beta}{4 \pi}\right)^2 c^2 -1 =0$; ($ii$) fixing the
horizon radius $r_+$, the temperature goes to zero as $T\sim \frac{1+3
  \alpha^2 r_+^2}{4 \pi \alpha r_+} r_B^{-1}$, these solutions are
stable if $3 \alpha^2 r_+^2 >1$, see \cite{Brown94}. 

\item[(IV)] the extremal cases require special care \cite{Hawk95,Zas,Ghosh}
  and were not  studied in any detail in this paper. 

\end{description}

\section{Conclusions}

We have studied the thermodynamics of the
Reissner-Nordstr\"om-anti-de~Sitter black hole  in York's
formalism. In the grand canonical ensemble where the temperature and
the electrostatic potential are fixed at a boundary with finite radius,
we have found one or two black hole solutions depending on the
boundary conditions and the value of the cosmological constant. In
general when there are two solutions, one is stable, the one with
larger event horizon radius, and the other is an instanton. On the
other hand, the cases with a single solution can correspond either to a
stable or unstable solution. We have found that both high
values of the cosmological constant and low temperatures favor stable
solutions. 

\section*{Acknowledgments}

C.S.P. acknowledges a research grant from JNICT FMRH/BIC/1535/95.

\newpage

\begin{references}

\bibitem{Hartle}
J.~B. Hartle and S.~W. Hawking, Phys. Rev. D {\bf 13},  2188  (1976).

\bibitem{GibH}
G.~W. Gibbons and S.~W. Hawking, Phys. Rev. D {\bf 15},  2752  (1977).

\bibitem{Hawklect}
S.~W. Hawking,  in {\em General Relativity: An Einstein Centenary Survey},
  edited by S.~W. Hawking and W. Israel (Cambridge University Press, Cambridge,
  1979).

\bibitem{Hawk76}
S.~W. Hawking, Phys. Rev. D {\bf 13},  191  (1976).

\bibitem{York}
J.~W. York, Phys. Rev. D {\bf 33},  2091  (1986).

\bibitem{WhitYork}
B.~F. Whiting and J.~W. York, Phys. Rev. Lett. {\bf 61},  1336  (1988).

\bibitem{Whit90}
B.~F. Whiting, Class. Quantum Grav. {\bf 7},  15  (1990).

\bibitem{grqcbrown}
J.~D. Brown and J.~W. York,  in {\em The Black Hole 25 Years After}, edited by
  C. Teitelboim and J. Zanelli (Plenum, New York, in press), gr-qc/9405024.

\bibitem{jolien}
J. D. E. Creighton, ``Gravitational Calorimetry'', PhD. thesis,
(Waterloo, 1996), gr-qc/9610038. 

\bibitem{Brown90}
J.~D. Brown {\it et~al.}, Class. Quantum Grav. {\bf 7},  1433  (1990).

\bibitem{Braden}
H.~W. Braden, J.~D. Brown, B.~F. Whiting, and J.~W. York, Phys. Rev. D {\bf
  42},  3376  (1990).

\bibitem{Brown94}
J.~D. Brown, J. Creighton, and R.~B. Mann, Phys. Rev. D {\bf 50},  6394
  (1994).

\bibitem{Louko}
J. Louko and S.~N. Winters-Hilt, Phys. Rev. D {\bf 54},  2647  (1996).

\bibitem{Lemos}
J.~P.~S. Lemos, Phys. Rev. D {\bf 54},  6206  (1996).

\bibitem{York89}
J.~W. York, Physica A {\bf 158},  425  (1989).

\bibitem{Callen}
H.~B. Callen, {\em Thermodynamics and an Introduction to Thermostatistics}
  (John Wiley \& Sons, New York, 1985).

\bibitem{HP83}
S.~W. Hawking and D.~N. Page, Commun. Math. Phys. {\bf 87},  577  (1983).

\bibitem{Page}
D.~N. Page,  in {\em Black Hole Physics}, edited by V.~D. Sabbata and Z. Zhang
  (Kluwer Academic, Dordrecht, 1992).

\bibitem{Hawk75}
S.~W. Hawking, Commun. Math. Phys. {\bf 43},  199  (1975).

\bibitem{Bek73}
J.~D. Bekenstein, Phys. Rev. D {\bf 7},  2333  (1973).

\bibitem{GB79}
G.~W. Gibbons and S.~W. Hawking, Commun. Math. Phys. {\bf 66},  291  (1979).

\bibitem{Yaffe}
D.~J. Gross, M.~J. Perry and L.~G. Yaffe, Phys. Rev. D {\bf 25},  330  (1982).

\bibitem{Hawk95}
S.~W. Hawking, G.~T. Horowitz, and S.~F. Ross, Phys. Rev. D {\bf 51},  4302
  (1995).

\bibitem{Zas}
O.~B. Zaslavskii, Phys. Rev. Lett. {\bf 76},  2211  (1996).

\bibitem{Ghosh}
A. Ghosh and P. Mitra, Phys. Rev. Lett. {\bf 78}, 1858 (1997).

\end{references}
\end{document}